\documentclass[pre,showpacs,amsmath]{revtex4}  

\newcommand{\beq}{\begin{equation}}
\newcommand{\enq}{\end{equation}}

\usepackage{graphicx}
\usepackage{amssymb}

\newcommand{\e}{\epsilon}

\renewcommand{\e}{\varepsilon}
\renewcommand{\epsilon}{\varepsilon}

\newcommand{\bi}{{\mathbf i}}
\newcommand{\sL}{{\mathcal L}}
\newcommand{\bj}{{\mathbf j}}
\renewcommand{\e}{\varepsilon}

\begin{document}
\title{
Multiscale dynamics of biological cells with chemotactic
interactions: from a discrete stochastic model to a continuous
description }

\author{Mark Alber$^{1*}$, Nan Chen$^1$, Tilmann Glimm$^2$ and Pavel M. Lushnikov$^{1,3}$
}

\affiliation{$^1$Department of Mathematics, University of Notre
Dame, Notre Dame,
46656\\
$^2$Department of Mathematics, Western Washington University,  Bellingham, WA 98225-9063\\
  $^3$ Landau Institute for Theoretical Physics, Kosygin St. 2,
  Moscow, 119334, Russia
}

\email{malber@nd.edu}

\date{
\today
}

\begin{abstract}
The Cellular Potts Model (CPM) has been used for simulating
various biological phenomena such as differential adhesion,
fruiting body formation of the slime mold \emph{Dictyostelium
discoideum}, angiogenesis, cancer invasion, chondrogenesis in
embryonic vertebrate limbs, and many others. In this paper, we
derive  continuous limit of discrete one dimensional CPM with the
chemotactic interactions between cells in the form of a
Fokker-Planck equation for the evolution of the cell probability
density function. This equation is then reduced to the classical
macroscopic Keller-Segel model. In particular, all coefficients
of the Keller-Segel model are obtained from parameters of the
CPM. Theoretical results are verified numerically by comparing
Monte Carlo simulations for the CPM with numerics for the
Keller-Segel model.

 \end{abstract}


\maketitle

* author for correspondence: Mark Alber

\section{Introduction \label{introduction}}
Biological cell dynamics has been studied at two main scales of
description. The macroscopic level provides one with a
coarse-grained treatment of biological cells through their
macroscopically averaged quantities such as local density of
cells
\cite{KellerSegel1970,BrennerLevitovBudrene1998,BrennerConstantinKadanoff1999,ErbanOthmer2004}.
The macroscopic scale is large in comparison with the typical
size of a cell. Macroscopic models are usually continuous and
utilize families of differential or integro-differential
equations to describe ``fields" of interaction. A much more
detailed approach is needed at the second, microscopic level
which takes into account stochastic fluctuations of the shape of
each individual cell.

Discrete models  describe individual (microscopic) behaviors of
cells. They are often applied to microscale events where a small
number of elements can have a large (and stochastic) impact on a
system. For example, while many periodic growth patterns can be
modeled using continuous methods, patterns which depend
sensitively on interaction between cells and substrate are best
modeled with discrete methods. Simplest discrete models describe
cells as point-wise objects. Some bacteria are self-propelled
 and do
 not change considerably their shape during motion (e.g.  {\it E. Coli}  \cite{BrennerLevitovBudrene1998,
Ben-Jacob} and {\it M. xanthus} \cite{Kaiser,Sozinova} bacteria).
They can be successfully represented as point-wise objects
undergoing reorientation while moving
\cite{BrennerLevitovBudrene1998,NewmanGrima2004,Alberreview1}. In
contrast, some other bacteria (e.g. \emph{Dictyostelium
discoideum} \cite{WeijerScince2003}) experience essential random
fluctuations of their shapes  and need to be treated  as extended
objects of variable shapes.

One of the microscopic models dealing with differential adhesion
and shape fluctuations is a Cellular Potts Model (CPM) which is an
extension of the well known Potts Model from statistical mechanics
\cite{granerglazier,glaziergraner}. In this model each biological
cell is represented by a cluster of pixels (spins). The CPM has
been used to simulate various biological phenomena such as cell
sorting \cite{granerglazier,glaziergraner}, fruiting body
formation of the slime mold  \emph{Dictyostelium discoideum}
\cite{maree1,maree2}, angiogenesis \cite{merks}, cancer invasion
\cite{turner_cancer}, chondrogenesis in embryonic vertebrate limbs
\cite{chat, compucell2}, and many others. (Different applications
of the CPM have been reviewed in \cite{Alberreview}.) Recently a
new alternative model was suggested
 \cite{NewmanMathBioEng2005}  which represents a cell as collection of subcellular
elements which interact with each other through phenomenological
intra- and intercellular potentials.

In addition to short range cell-cell adhesion and interactions
between cells and their surrounding extracellular matrix
(haptotaxis), cell interact at long range through signal
transmission and reception mediated by a diffusing chemical field
(chemotaxis). Continuous macroscopic Keller-Segel model of the
evolution of the density of cells with chemotactic interactions
have been extensively studied
\cite{KellerSegel1970,BrennerLevitovBudrene1998,
BrennerConstantinKadanoff1999,ErbanOthmer2004} over the years. In
particular, it has been successfully applied to the description of
{\it Escherichia coli} bacteria aggregation due to chemotaxis in
\cite{BrennerLevitovBudrene1998}. The drawback of continuous
models is that they have a lower resolution than discrete models.
However, their  advantage is the availability of a large set of
analytical and numerical tools
 for analyzing solutions of
the corresponding nonlinear partial differential equations (PDEs).
By contrast, the analytical study of discrete models is often
impossibly complicated, and their computational implementation is
often much less efficient in comparison with numerical methods
available for  PDEs. It is thus important, for numerical,
analytical, as well as conceptual reasons to establish connections
between various discrete and continuous models of the same
biological problem.

There is a vast literature on studying continuous limits of
point-wise discrete microscopic models. In particular, classical
Keller-Segel model has been derived from a model with point-wise
representation for cells undergoing random walk
\cite{Alt1980,StevensSIAM2000,NewmanGrima2004}. However, much less
work has been done on deriving macroscopic limits of microscopic
models which treat cells as extended objects. One of the first
attempts at combining microscopic and macroscopic levels of
description of cellular dynamics has been described in
\cite{turner2004} where the diffusion coefficient for a collection
of noninteracting randomly moving cells has been derived  from a
one dimensional CPM. Recently a microscopic limit of subcellular
elements model \cite{NewmanMathBioEng2005} was derived in the form
of continuous advection-diffusion equation for cellular density.
In the present paper,  we establish a connection between a
one-dimensional CPM of a cell moving on a substrate and reacting
to a chemical field, and a Fokker-Planck equation for the cell
probability density function. This equation is then reduced to the
classical macroscopic Keller-Segel equation. In particular, we
derive all coefficients of the Keller-Segel model from parameters
of the CPM. We also compare Monte Carlo simulations for the CPM
with numerics for the Keller-Segel model to support our
theoretical results.

\begin{figure}
\centering
    \includegraphics[width=3.5in]{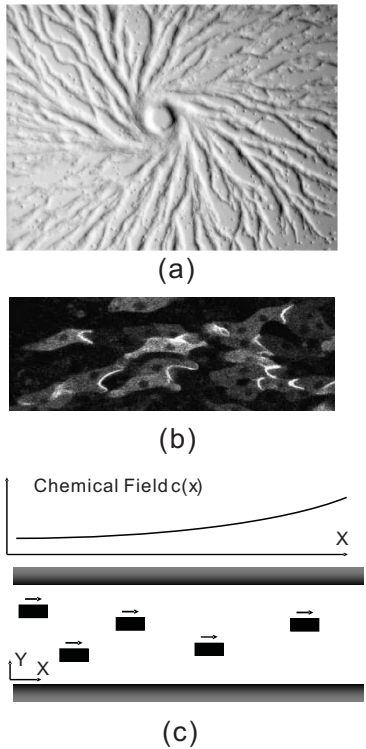}
    \caption{(a) Streaming of {\it Dictyostelium discoideum} towards
the aggregation center. Cells move chemotactically towards the
aggregation center leading to formation of cell streams and
finally mounds. (Reproduced from \cite{dundee} with permission).
(b) Example of a quasi-one-dimensional motion of {\it
Dictyostelium discoideum} inside a stream (this picture is on much
smaller scale compared with (a)). Cells are moving parallel to
each other in the  direction of chemical gradient (from left to
right). Chemical gradient also causes polarization of cells so
that they become elongated in the direction of a gradient.
(Reproduced from \cite{WeijerScince2003} with permission.)  (c)
Schematic picture of cell motion in a gradient of chemical field
(e.g. chemo-attractant cAMP). The concentration of the chemical field is
shown schematically above the main figure. }\label{fig:fig2}
\end{figure}

Unified multiscale approach, described in this paper and based on
combining microscopic and macroscopic models, can be applied to
studying such biological phenomena as streaming in
\emph{Dictyostelium discoideum}. In starved populations of {\it
Dictyostelium} amoebae, cells produce and detect a communication
chemical (cAMP). The movement of \emph{Dictyostelium} cells
changes from a random walk to a directed walk up the cAMP gradient
resulting in formation of streams of cells towards the aggregation
center (see Fig. \ref{fig:fig2}a) and subsequent formation of
multi-cellular fruiting body. Figure \ref{fig:fig2}b shows cells'
movement from left to right in response to waves of cAMP
travelling through the aggregation stream from right to left. The
cAMP gradient on the up-down direction is very small and could be
ignored. Figure \ref{fig:fig2}c schematically demonstrates the
main features of the cell movement.
 Unlike  differential adhesion
\cite{granerglazier,glaziergraner}, chemotactic cell motion is
highly organized over a length scale significantly larger than the
size of a single cell. (For details about modeling {\it
Dictyostelium discoideum} fruiting body formation see e.g.
\cite{maree1,maree2,DormannVasievWeijer2002,VasievHogewegPanfilov1994}).

The paper is organized as follows. In Section \ref{descrCPM} we
describe a one dimensional CPM with chemotaxis. In Section
\ref{master}, we derive from the Monte Carlo dynamics of the CPM
the  discrete master equation for the probability density function
$P(x,L,t)$; that is, the probability that at time $t$, there is a
cell whose length is $L$ and whose center of mass is located at
$x$. In Section \ref{continuousCPM} we use the discrete master
equation to derive a partial differential equation for $P(x,L,t)$
in a continuous limit which assumes that cell changes its
position and length at each Monte Carlo step by a small amount.
We show that the dependence of $P(x,L,t)$ on $L$ is very close to
the Boltzmann distribution. This is used in Section
\ref{FokkerPlanck} for the derivation of a Fokker-Planck equation
for the probability density function $p(x,t)$  of a cell's center
of mass being at $x$ which is the main result of the paper. In
Section \ref{KellerSegel} it is shown that addition of the time
dependence of chemical field reduces the Fokker-Planck equation
to the Keller-Segel equations. Section \ref{numerics} deals with
numerical verification of the theoretical results of the previous
sections and compares the Monte Carlo simulations for our CPM and
Keller-Segel models.

\section{The Cellular Potts Model\label{descrCPM}}

The Cellular Potts Model (CPM), an extension of the Potts Model from statistical mechanics,
is a flexible and powerful way to model cellular patterns.
Its core mechanism is the competition between the minimization of various
energy terms in some generalized functional of the cellular configuration, e.g., surface
minimization, cell-cell contact and chemotactic interactions, and global geometric
constraints. It simulates stochastic fluctuations of
cell shapes as simple thermal fluctuations.

The CPM is defined on a rectangular lattice $\sL$, which is of the
form $[0,m_x]$ (for 1 dimension), $[0,m_x]\times [0,m_y]$ (for 2
dimensions) or  $[0,m_x]\times [0,m_y]\times [0,m_z]$ (for 3
dimensions). (Here $[0,m]=\{0,1,\ldots,m\}$.) The elements of
$\sL$ are called the {\em lattice sites} (intervals in 1D, pixels
in 2D, voxels in 3D). A lattice site is denoted by a index $\bi\in
\sL$.

Each lattice site has an assigned ``spin" $\sigma(\bi)$ which can
have values $s=0,1,\ldots, Q$, where $s=0$ corresponds to absence
of any cell at the given site and the value $1\le s\le Q$ means that
the given site is occupied by the $s$th cell, where $Q$ is the total
number of cell in the system. Assume that we fix the values of
$\sigma(\bi)$ at each lattice site, then we refer to that set of
values as a {\em configuration}. The best way to visualize a
configuration is to regard the different spins as different
colors. Each lattice site $\bi$ has a color $\sigma(\bi)$. The
cells are the collections of lattice sites that have the same
spin (color), so that each lattice site can be occupied by a
single cell only. White color correspond to absence of any cell at
given site: $\sigma(\bi)=0.$ In the model considered here we
assume that cells cannot divide so that sites with the same
color are always connected.

We assume periodic boundary conditions so that pixels at zero
position in $x,y $ or $z$ are identical to sites with $i_x=m_x+1,
\ i_y=m_y+1$ and $i_z=m_z+1,$ respectively.

The temporal dynamics of the system is defined by certain
probabilistic transition rules between the configurations, giving
rise to a Markov chain of configurations, i.e. a sequence of
configurations $\sigma^0, \sigma^1, \sigma^2,...$. To describe
the transition rules, we associate to each configuration $\sigma$
an energy $E(\sigma)$, also referred to as the Hamiltonian. The
state changes from one configuration to the next are governed by
an energy minimization principle with effective temperature $T.$
This is implemented by means of the Metropolis algorithm for
Monte-Carlo Boltzmann dynamics. The algorithm works as follows:

Given a configuration $\sigma^n$, we randomly select a lattice
site $\bi\in\sL$ such that not all of its nearest lattice
neighbors have  the same spin. We then randomly choose a lattice
neighbor $\bi'$ of $\bi$ with $\sigma^n(\bi')\neq\sigma^n(\bi)$.
Let $\sigma'$ be the configuration we obtain by ``flipping'' the
spin of $\bi$, i.e. we have $\sigma'(\bj)=\sigma^n(\bj)$ for all
$\bj\neq \bi$, and $\sigma'(\bi)=\sigma^n(\bi')$. The new
configuration $\sigma^{n+1}$ is then either $\sigma^n$ or the
configuration $\sigma'$. The probability $\Phi(\Delta E)$ that
$\sigma'$ is accepted as the next configuration $\sigma^{n+1}$
depends on the energy difference $\Delta
E=E(\sigma')-E(\sigma^n)$. The formula is
\begin{equation}\label{Phi}
        \Phi(\Delta E)=\begin{cases} 1,\quad & \text{if } \Delta E\leq 0\\
                                \exp(-\beta\Delta E), &  \text{if } \Delta E> 0
                        \end{cases}.
\end{equation}
Here $\beta=1/T$ is a positive constant.

\begin{figure}
\includegraphics[width=0.5\textwidth]{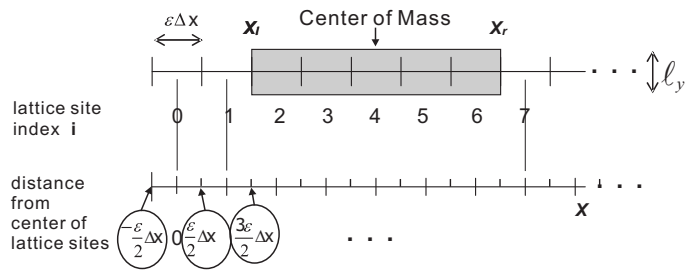}
\caption{Example of a cell in one-dimensional CPM. The cell
(shaded domain) occupies lattice sites $2,\ldots,6$. It has a
length of $5\e\Delta x$, its center of mass is located at
$x=4\e\Delta x$, and its end points are $x_l=1.5\e\Delta x$ and
$x_r=6.5\e\Delta x$. }\label{fig:fig1}
\end{figure}

In this paper, we consider a quasi-one-dimensional CPM, which
means that cells are assumed to move along $x$ direction only and
have fixed thickness $l_y$ in the $y-$direction (see Fig.
\ref{fig:fig1}). Let $\epsilon\Delta x$ denote  the size of
lattice site, where $0<\e\ll 1$, $\e$ is the small dimensionless
constant and $\triangle x$ is a dimensional constant of the order
of one.  Each lattice site is described by its index
$\bi=0,1,\ldots,$ so that the center of each lattice site is
located at $x=\bi\e\Delta x$ with the lattice site left border at
$x_l=(\bi-\frac{1}{2})\e\Delta x$ and the lattice site right
border at $x_r=(\bi+\frac{1}{2})\e\Delta x$  (see
Figure~\ref{fig:fig1}.)

In what follows, we will consider the dynamics of a single cell so
that the spin $\sigma$ can take two values: 0 if cell is absent at
a given site and 1 if cell occupies a given site. However, our results
remain valid for an ensemble of $n$ cells which are well
separated from each other, so that the probability that two cells
would try to occupy the same volume is negligible. This allows us
to neglect cell-cell contact interactions. We assume that cells can
interact only with the substrate (haptotaxis)  and the chemical
field $c(x)$ (chemotaxis). The chemical field is assumed to depends
only on $x$ but not on $y$. Cells can also produce a chemical
which then diffuses. In Section \ref{KellerSegel} we discuss
production of chemicals by cells.

A natural biological realization of this quasi-one-dimensional
model is the motion of biological cells in
streams\cite{VasievHogewegPanfilov1994}. E.g. the amoebae {\it
Dictyostelium discoideum} under starving condition typically
forms streams\cite{DormannVasievWeijer2002}. The biological cells
inside each stream are moving towards the aggregation center (see Fig.
\ref{fig:fig2}a), which results in complicated 2D patterns
\cite{VasievHogewegPanfilov1994}. If we zoom to a small scale,
we will see that the motion of cells inside each stream is
quasi-one-dimensional with cells moving parallel to each other in
$x-$direction ( Fig. \ref{fig:fig2}b).  The chemical gradient of
the other direction ($y$ direction) could be neglected and
 during cells movement there is no cell-cell interactions, such as cell collisions or cell
 signaling. Fig. \ref{fig:fig2}c
schematically shows such a parallel motion of the cells from left
to the right under the action of the gradient of a chemical field
(chemo-attractant).

For a given configuration $\sigma$ of spins, let $N=N(\sigma)$
denote the number of lattice sites that the cell occupies. The
length of the cell is equal to $L=N\e\Delta x$. We denote the
position of the center of mass of the cell by $x$ and denote the
position of the left and right ends of the cell by $x_l$ and
$x_r$, respectively.  Then $L=x_r-x_l$. (See Figure~\ref{fig:fig1}.)

We assume that the chemical field $c(x)$ is a slow function of
time so its typical time scale is much bigger than the time step
of a Monte Carlo algorithm. Then the Hamiltonian is given by the
formula:
\begin{equation}\label{Hamiltonian}
        E=J_{cm}\cdot(2L+2\ell_y)+\lambda \,(L-L_T)^2+\mu\, c(x)\, L.
\end{equation}
The first term is a surface energy term which corresponds to the
cell-substrate interaction energy (haptotaxis), where $J_{cm}$ is
an interaction energy between the cell and the medium per unit
length.  The second term is a length-constraint term which
penalizes deviations of the cell length $L$ from the target cell
length $L_T$. Here $\lambda$ is a positive constant. The choice of
$\lambda$ and $\beta$ is determined by the typical scale of
fluctiuations of the cellular shape.  The third term in
(\ref{Hamiltonian}) is the coupling chemical energy. This term
will favor cell motion {\em down} or {\em up} the chemical
gradient for $\mu>0$ and $\mu<0$, respectively. We assume that the
concentration $c(x)$ is a slow function of $x$ on a scale of the
typical cell's length $L:$
\begin{equation}\label{xLcondition}
x_c/L\gg 1,
\end{equation}
where $x_c$ is a typical scale for variation of $c(x)$ in $x$.
This is consistent with the generally accepted view that cells are
typically too small to detect chemical gradients without moving.
(See e.g. \cite{Adler}; however recent experimental evidence may
put this view in question \cite{Thar}.) Note that the chemical energy
could also be defined as  $\mu \int_{x_l}^{x_r}c(x)dx$. But in the limit
$(\ref{xLcondition})$, this is equivalent to the form used in the
Hamiltonian (\ref{Hamiltonian}).

\section{Discrete evolution equation for probability density function \label{master}}
In this section, we develop an analytical model for the evolution
of the stochastic dynamics of a cell in CPM.

Let $P(x,L,t)$ be a probability density  for the cell with the
center of mass at $x$ of length $L$ at time $t$. Spins
$\sigma(\bi)$ are defined on the lattice $\sL$ so that the length
of the cell $L$, which is the difference between positions of
right and left ends of cell: $L=x_r-x_l,$ can take values
$n\epsilon\triangle x,$ $n=1,2,\ldots$. The position of the
center of mass $x=(x_r+x_l)/2$ can take values
$n\epsilon\triangle x/2,$ $n=1,2,\ldots$. That is, the CPM grid
is twice the size of the grid of center of mass. In particular,
if $2\frac{x}{\epsilon \triangle x}$ is an even number (i.e. $x$
coincides with one of the lattice sites) then the ratio
$\frac{L}{\epsilon \triangle x}$ is also an even number.
Alternatively, if $2\frac{x}{\epsilon \triangle x}$ is an odd
number (i.e. $x$ coincides with a boundary between two
neighboring lattice sites) then the ratio $\frac{L}{\epsilon
\triangle x}$ is an odd number.

For convenience, we choose a normalization for $P(x,L,t)$ such
that the probability for a cell to have its center of mass at $x$
and length $L$ at time $t$ is given by $(\epsilon\triangle
x)^2P(x,L,t)$. The factor $(\epsilon\triangle x)^2$ results from
the product of $\epsilon \triangle x/2$ (the spacing between
lattice sites) and  $2\epsilon \triangle x$ (the spacing in $L$
for a fixed $x$). With this normalization, $P(x,L,t)$ becomes a
true probability density  in the continuous limit $\epsilon \to
0.$

We choose the time interval between two Monte Carlo steps to be
$\e^2\Delta t$, where $\Delta t$ is a fixed constant of dimension
of time. This implies diffusive time-space scaling,
\[
        \frac{\e^2\Delta t}{(\e\Delta x)^2}=\frac{\Delta t}{(\Delta x)^2}
\]
which is independent of the scaling parameter $\e$. We now switch
from measuring time in Monte Carlo steps $n=0,1,\ldots$, to a
continuous time variable $t=n\,\e^2\,\Delta t$.

Suppose at time $t$ the cell is at a state $(x,L)$ meaning that it
has length $L$ and its center of mass is at $x$. The stochastic
discrete system at time $t+\epsilon^2\triangle t$ can switch to
one of the following four possible states:
\begin{enumerate}
\item[(a)] $(x+\epsilon\Delta x/2,L+\epsilon\triangle x)$ by adding the lattice site $x_r+\e\Delta x$ to the
right end of cell;
\item[(b)] $(x+\epsilon\Delta x/2,L-\epsilon\triangle x)$ by taking away the site
$x_l$ from the left end of the cell;
\item[(c)] $(x-\epsilon\Delta x/2,L+\epsilon\triangle x)$ by adding the lattice site $x_l+\e\Delta x$ to the
left end of cell;
\item[(d)] $(x-\epsilon\Delta x/2,L-\epsilon\triangle x)$ by taking away the site
$x_r$ from the right end of the cell.
\end{enumerate}
 Therefore, the most general  master equation for evolution of the probability density  $P(x,L,t)$ has the form
\begin{eqnarray}\label{pmasterxL1}
P(x,L,t+\epsilon^2\triangle t)=\big
[1-T_l(x-\frac{\epsilon}{2}\triangle x, \, L+\epsilon \triangle
x;\, x,L, t) -T_r(x+\frac{\epsilon}{2}\triangle x, \, L+\epsilon
\triangle x;\, x,L, t)\nonumber \\
-T_l(x+\frac{\epsilon}{2}\triangle x, \, L-\epsilon \triangle
x;\, x,L, t)-T_r(x-\frac{\epsilon}{2}\triangle x, \, L-\epsilon
\triangle x;\, x,L, t)\big ]P(x,L,t)
\nonumber \\
+T_l(x,L; \, x+\frac{\epsilon}{2}\triangle x,\, L-\epsilon
\triangle x,t)P(x+\frac{\epsilon}{2}\triangle x,\, L-\epsilon
\triangle x,t)
\nonumber \\
+T_r(x,L; \, x-\frac{\epsilon}{2}\triangle x,\, L-\epsilon
\triangle x,t)P(x-\frac{\epsilon}{2}\triangle x,\, L-\epsilon
\triangle x,t)
\nonumber \\
+T_l(x,L; \, x-\frac{\epsilon}{2}\triangle x,\, L+\epsilon
\triangle x,t)P(x-\frac{\epsilon}{2}\triangle x,\, L+\epsilon
\triangle x,t)
\nonumber \\
+T_r(x,L; \, x+\frac{\epsilon}{2}\triangle x,\, L+\epsilon
\triangle x,t)P(x+\frac{\epsilon}{2}\triangle x,\, L+\epsilon
\triangle x,t) ,
\end{eqnarray}
where $T_l(x,L;x',L')$ and $T_r(x,L;x',L')$ correspond to
transitional probabilities for a cell of length $L'$ and center of
mass at $x'$ to change into a cell of length $L$ and center of
mass at $x'$. Subscripts ``l" and ``r" corresponds to transition
due to addition/removal of a pixel from the left/right side of a
cell respectively. These transition probabilities are given by
\begin{eqnarray}\label{Tdef1}
T_l(x,L;x',L')=T_r(x,L;x',L')= \frac{1}{4}\Phi\Big (E(x,L)-
E(x',L' ) \Big ),
\end{eqnarray}
where $E(x,L)$ is the Hamiltonian $(\ref{Hamiltonian})$ and
$\Phi(\triangle E)$ is given by Eq. $(\ref{Phi})$. Factor $1/4$ in
$(\ref{Tdef1})$ accounts transitions  to 4 possible states
(a)-(d). For computational purposes it is convenient to rewrite
$(\ref{Phi})$ in an equivalent form
\begin{eqnarray}\label{Phidef}
\Phi(\triangle E)=1-\Big \{1-\exp\big [-\beta\triangle E\, \big
]\Big \} \Theta(\triangle E).
\end{eqnarray}
Here $\Theta(x)$ is a Heaviside step function: $\Theta(x)=1$ for
$x> 0$ and $\Theta(x)=0$ for $x<0$.

\section{Continuous evolution equation for probability density function of CPM \label{continuousCPM}}

Below we assume $\epsilon$ to be small, $\epsilon\ll 1,$ so that
the change of the cell size and position is small at each
Monte-Carlo step. Now we carry out a Taylor series expansion in
$\epsilon$ of the terms in Eq. $(\ref{pmasterxL1})$. One has to
take special care of $\Theta(\triangle E)$ terms in the expansion
because the Heavyside step function is not analytic. To avoid
this difficulty we do not expand the function itself but only its
argument instead. There is an important simplification which
comes from the fact that $\Theta(\triangle E)+\Theta(-\triangle
E)=1$ so that in Eq. $(\ref{pmasterxL1})$ we obtain that
$T_{l,r}(x,L;\, x',L', t)+T_{l,r}(x',L'; \,x,L, t)=(1/4)\exp\big
(-\beta|E(x,L)-E(x',L')|\big )$. This yields mutual cancellation
of nonanalytical terms up to order $O(\epsilon^2)$. Then,
equating coefficients in the Taylor expansion in Eq.
$(\ref{pmasterxL1})$ in order $O(\epsilon^2)$ results in the
Fokker-Planck equation
\begin{eqnarray}\label{pottscontinuousfull1}
\partial _t P(x,L,t)=D(\partial^2_x
+4\partial^2_L)P+8D\beta\lambda\partial_L(\tilde L P)+D\beta
L\mu\partial_x\big [c'(x)P)\big ], \nonumber \\
\tilde L=\frac{1}{\lambda}\big [J_{cm}+\lambda (L-L_T)
+\frac{1}{2}\mu c(x) \big ], \quad D=\frac{(\triangle x)^2}{8
\triangle t}.
\end{eqnarray}

Now, under certain conditions to be described in the end of this
section, the terms
$D4\partial^2_LP+8D\beta\lambda\partial_L(\tilde L P)$ dominate
the other terms on the right hand side of Eq.
$(\ref{pottscontinuousfull1})$. This means that at the leading
order, one can neglect terms with $x-$derivatives. Under this
assumption, the probability density function $P(x,L,t)$
approaches a Boltzmann distribution for cell length exponentially
in time at the rate of $8D\beta\lambda$:
\begin{eqnarray}\label{Pinitial}
P(x,L,t)=P_{Boltz}(x,L)p(x,t),
\end{eqnarray}
where $p(x,t)$ is a probability density function of finding cell's
center of mass at $x$.  $P_{Boltz}(x,L)$ is the Boltzmann
distribution for the cell length given by
\begin{eqnarray}\label{Boltzmann}
P_{Boltz}(x,L)=\frac{1}{Z}\exp(-\beta \triangle E_{length}), \\
\label{dElengthdef} \triangle
E_{length}=E(L)-E_{min}=\lambda\tilde L^2,
\end{eqnarray}
where $E_{min}$ is a minimum of energy $E(L)$ as a function of $L$
for a given $x$,
\begin{eqnarray}\label{Emin}
E_{min}=E(L_{min}), \quad
L_{min}=L_T-\frac{J_{cm}}{\lambda}-\frac{\mu c(x)}{2\lambda},
\end{eqnarray}
and $Z$ is a partition function
\begin{eqnarray}\label{Zpart}
Z(x)=2\epsilon\triangle x\times \sum\limits_{L=(1+\alpha)\epsilon
\triangle x, \, (3+\alpha)\epsilon \triangle x, \,
(5+\alpha)\epsilon \triangle x,\ldots}\exp(-\beta \triangle
E_{length}),
\nonumber \\
\alpha=1 \ \mbox{for} \ \frac{x}{\epsilon \triangle x}=n, \quad
%
%
\alpha=0 \ \mbox{for} \ \frac{x}{\epsilon \triangle x}=n+1/2,
\quad n\in\mathbb{N}.
\end{eqnarray}
Here we use the fact that due to discrete nature of our model, the
position of the center of mass, $x$, could be located at one of
the lattice sites $x=m\epsilon \triangle x$ ($m$ being an integer
number) if the length of the cell $L$ is an even number of units
${\epsilon \triangle x}$ or $x$ could be located at the boundary
between two neighboring lattice sites in case of $L$ being equal
to an odd number of units of ${\epsilon \triangle x}$. The factor
$(\epsilon\triangle x)^2$ in the definition of the partition
function $(\ref{Zpart})$ is chosen in such a way as to yield $\int
P(x,L,t)dLdx=1$ in the continuous limit. We can also normalize
$\int P(x,L,t)dLdx=N$ to the total number of cells in the system
$N$.

In the continuous limit, $\epsilon \to 0,$ the sum in Eq.
$(\ref{Zpart} )$ is transformed into the integral
\begin{eqnarray}\label{Zpartcont}
Z\simeq\int^{+\infty}_{-\infty}\exp(-\beta \triangle
E_{length})dL=\frac{\sqrt{\pi}}{\sqrt{\beta \lambda}}, \quad x\to
0.
\end{eqnarray}
Here we have extended the limits of integration from $(0,+\infty)$ to
$(-\infty,+\infty)$. Of course physically, the length of the cell $L$ is
always positive. A typical fluctuation of the cell size $\delta
L=L-L_{min}$ about $L_{min}$ is determined by the Boltzmann
distribution $(\ref{Pinitial})$ as $\beta\lambda \delta L^2\sim
1$. In what follows we make a biologically motivated assumption
about fluctuations of the cell size being much smaller than $L$:
$|\delta L|\ll L_{min}$ which results in the condition
\begin{eqnarray}\label{betacond0}
\beta  L_{min}^2\lambda \gg 1.
\end{eqnarray}
This justifies the use of the integration limits $(-\infty,+\infty)$ in
Eq. $(\ref{Zpartcont})$ instead of $(0,+\infty)$ because under
this condition $\exp(-\beta \triangle E_{length})$ peaks around
$L_{min}$ and replacement of integration limits results in an
exponentially small correction.

Let us now specify the conditions for the applicability of the Boltzmann
distribution approximation $(\ref{Pinitial})$.
For this, consider Eq. $(\ref{pottscontinuousfull1})$. We have
$\beta\lambda \delta L^2\sim 1$. We now assume in addition the relation
\begin{eqnarray}\label{betacond1}
\beta  x_0^2\lambda \gg 1,
\end{eqnarray}
where $x_0$ is a typical scale of $P$ with respect of $x$. Note that under
the assumption that $L_{min}\ll x_0$, i.e. that the typical length of a
cell is much smaller than $x_0$, the condition
$(\ref{betacond1})$ follows from $(\ref{betacond0})$.
It follows from $(\ref{betacond1})$ that $|\partial^2_xP|\ll|4\partial^2_LP|$, and consequently, we may
neglect the first term with $x-$derivative , $\partial_{xx}P$, on the right hand side of Eq.
$(\ref{pottscontinuousfull1})$.

The second condition for the applicability of the Boltzman
distribution approximation (\ref{Pinitial}) is the assumption
that the last term with $x-$derivative in Eq.
$(\ref{pottscontinuousfull1})$ is small, $\big |\beta
L\mu\partial_x\big [c'(x)P\big ]\big |\ll |4\partial^2_LP|$. This
is true if
\begin{eqnarray}\label{lambcond1}
|L_{min}\mu c_0| \big (1+\frac{x_c}{x_0}\big ) \ll \lambda x_0^2,
\end{eqnarray}
where $c_0$ is a typical amplitude of $c(x)$ and $x_c$ is a
typical scale of variation of $c(x)$ with respect to $x$. Lastly,
recall that we derive the continuous Eq.
$(\ref{pottscontinuousfull1})$  from the master equation
$(\ref{pmasterxL1})$ under the condition of the step in $x$ being
small
\begin{eqnarray}\label{epscond1}
\epsilon \ll 1.
\end{eqnarray}
Notice that diffusion coefficient $D$ in Eq.
$(\ref{pottscontinuousfull1})$ does not depend on $\beta$.
Instead $\beta$ determines a rate of convergence
$\tau_r^{-1}=8D\beta\lambda$ of $P(x,L,t)$ to the Boltzmann
distribution $(\ref{Pinitial})$.

We have solved both the master equation $(\ref{pmasterxL1})$ and
its continuous limit $(\ref{pottscontinuousfull1})$ numerically
with initial conditions $P(x,L,0)$ different from the Boltzmann
distribution $(\ref{pmasterxL1})$. Simulations described in Section
\ref{numerics} demonstrate that for each $x$, the solution $P(x,L,t)$
indeed converges in time to the Boltzmann
distribution at an exponential rate of $\sim 8D\beta\lambda$.

\section{Fokker-Planck equation for probability density function $p(x,t)$ \label{FokkerPlanck}}

We now  turn to calculating the probability  density function
$p(x,t)$ of a center of cell's mass being at $x$. It is given by
the sum over all possible lengths of a cell
\begin{eqnarray}\label{pxdef3}
p(x,t)=2\epsilon\triangle x\sum\limits_{L=(1+\alpha)\epsilon
\triangle x, \, (3+\alpha)\epsilon \triangle x, \,
(5+\alpha)\epsilon \triangle x,\ldots}P(x,L,t)\simeq
\int^{+\infty}_{-\infty}
P(x,L,t)dL, \quad \epsilon\to 0, \nonumber \\ %
\alpha=1 \ \mbox{for} \ \frac{x}{\epsilon \triangle x}=n, \quad
%
%
\alpha=0 \ \mbox{for} \ \frac{x}{\epsilon \triangle x}=n+1/2,
\quad n\in\mathbb{N},
\end{eqnarray}
which reduces to Eq. $(\ref{Pinitial})$ in the Boltzmann
distribution approximation limit.

To derive closed equation for $p(x,t)$ we substitute ansatz
$(\ref{Pinitial})$ into $(\ref{pottscontinuousfull1})$ and
integrate both right hand and left hand sides of Eq.
$(\ref{pottscontinuousfull1})$ with respect to $L$ to obtain
\begin{eqnarray}\label{pottscontinuous2}
\partial _tp=D\partial^2_x p-\partial_x\big [\chi(x)p\, \partial_x c(x)\big ], \nonumber \\
\chi(x)= \frac{D}{\lambda}\beta\mu  \Big [J_{cm}-\lambda L_T
+\frac{1}{2}\mu c(x)\Big ],\quad D=\frac{(\triangle x)^2}{8
\triangle t}.
\end{eqnarray}
This continuous equation is the main result of this paper.
The conditions for the applicability of Eq. $(\ref{pottscontinuous2})$
are given by Eqs. $(\ref{betacond0})$, $(\ref{betacond1})$,
$(\ref{lambcond1})$ and $(\ref{epscond1})$.

\section{Reduction to Keller-Segel model \label{KellerSegel}}

In this section we add time dependence to the chemical field $c$
(concentration of chemoattractant or chemorepellant) by including
a diffusion equation with the source term $ap$ which determines
the secretion of chemical by a cell
\begin{eqnarray}\label{ceq1}
\partial_tc=D_c\partial^2_xc-\gamma c +a\, p,
\end{eqnarray}
where $D_c$ is a diffusion coefficient of the chemical field,
$\gamma$ is the decay rate of the chemical field and $a$ is a
production rate of the chemical field.

The system of equations $(\ref{pottscontinuous2})$ and
$(\ref{ceq1})$ is applicable  under the assumption that the
typical time scale $\tau_c$ of diffusion of $c(x,t)$, given by
$\tau_c=\frac{\triangle x_c^2}{D_c}$, is large in comparison with
convergence time $\tau_r=1/(8 D\beta\lambda)$ of $P(x,L,t)$ to the
Boltzmann distribution $(\ref{Pinitial})$, where $x_c$ is a
typical spacial width of the distribution of $c(x,t)$. Namely,
this condition has the form
\begin{eqnarray}\label{taucond2}
\tau_c/\tau_r=8D\beta\lambda\tau_c\gg 1,
\end{eqnarray}

Eqs. $(\ref{pottscontinuous2})$ and $(\ref{ceq1})$ form a closed
set of equations which is equivalent to the classical Keller-Segel
model \cite{KellerSegel1970} of chemotaxis. If the parameters satisfy
condition
\begin{eqnarray}\label{KellerSegelcond1} |J_{cm}-\lambda L_T|\gg
\frac{1}{2}|\mu| c(x),
\end{eqnarray}
than Eq. $(\ref{pottscontinuous2})$ reduces to the following
commonly used form of the Keller-Segel model
\cite{BrennerConstantinKadanoff1999,BrennerLevitovBudrene1998}:
\begin{eqnarray}\label{pottscontinuousKellerSegel}
\partial _tp=D\partial^2_x p-\chi_0\partial_x\big [p\, \partial_x c\big ], \nonumber \\
\chi_0= D{\lambda}\beta\mu  \Big [J_{cm}-\lambda L_T \Big ],\quad
D=\frac{(\triangle x)^2}{8 \triangle t}.
\end{eqnarray}
The probability density function $p(x,t)$ corresponds to the
microscopic density in the Keller-Segel model. Notice that both
in the Keller-Segel model and CPM  considered in this paper,
there is no direct interaction between cells except through
production and reaction to a chemoattractant. In other words,
cells are treated in a way similar to a dilute gas with long
range nonlocal interactions due to reaction to a chemical field.

\section{Comparison of numerical simulations\label{numerics}}

In this section, we describe numerical tests  comparing Monte
Carlo simulations of the CPM and simulations of both discrete and
continuous models for the probability density functions $P(x,L,t)$
and $p(x,t)$, as given by Eqs.  $(\ref{pmasterxL1})$,
$(\ref{pottscontinuousfull1})$ and $(\ref{pottscontinuous2})$.

\subsection{Monte Carlo simulations \label{MT}}
The computation of the frequency distribution of the cell center
of mass and length for the  CPM has been carried out as follows:
\begin{enumerate}
\item We run a large number $N$ of CPM simulations with one cell
with the same initial conditions. \item We fix a time interval
$\delta t=\epsilon^2 \triangle t$, i.e. we fix the time interval
between successive Monte Carlo steps. For each simulation we
record the locations of the center of mass and and lengths of the
cell at the times $t=\delta t, 2\delta t, 3\delta t,\ldots$
\item After the $N$ runs, the recorded data give a frequency
distribution $M(x,L,t)$ for the location of the center of mass of
the cell and length of the cell.
\end{enumerate}

The frequency distribution $M(x,L,t)$ determines the approximation
$P_{\text{cpm}}(x,L,t)=M(x,L,t)/(N(\epsilon \triangle x)^2)$  of
the probability density function $P(x,L,t)$ for the center of
mass of a cell of length $L$ being at $x$ at time $t$. Therefore,
we compare $P_{\text{cpm}}(x,L,t)$  with $P(x,L,t)$ which is a
solution of  either the master equation $(\ref{pmasterxL1})$ or
the Fokker-Planck equation $(\ref{pottscontinuousfull1})$. To
approximate the probability density function of center of mass
$p(x,t)$ we sum up over all values of $L$ on the grid in a way
used in Eq. $(\ref{pxdef3})$
\begin{eqnarray}\label{pxpottsdef3}
p_{cpm}(x,t)=2\epsilon\triangle x\sum\limits_{L=(1+\alpha)\epsilon
\triangle x, \, (3+\alpha)\epsilon \triangle x, \,
(5+\alpha)\epsilon \triangle
x,\ldots}P_{cpm}(x,L,t), \nonumber \\ %
\alpha=1 \ \mbox{for} \ \frac{x}{\epsilon \triangle x}=n, \quad
%
%
\alpha=0 \ \mbox{for} \ \frac{x}{\epsilon \triangle x}=n+1/2,
\quad n\in\mathbb{N}.
\end{eqnarray}
In what follows, we compare $p_{\text{cpm}}(x,t)$ for $\epsilon\ll
1$ with $p(x,t)$, a solution of the continuous Eq.
$(\ref{pottscontinuous2})$, corresponding to the following choice
of parameters
\begin{align}\label{default_param}
  \lambda=4, L_T=5, J_{cm}=2, \beta=15, \mu=0.1, \Delta x=1, \Delta
  t=1.
\end{align}
The size of the CPM lattice is chosen to be $L_{cpm}=100$; and the
model is typically run from $t_0=0$ to $t_{end}=200$. The number
of the CPM lattice sites and the number of Monte Carlo steps are
chosen to be $\frac{L_{cpm}}{\e\Delta x}$ and $
\frac{t_{end}}{\e^2\Delta t}$ respectively. We use a range of
values of $\epsilon$ between $0.2$ and $0.001$.

The initial conditions for each CPM run are chosen as follows. A
random pixel in the interval $[40, 60]$ is selected as a center of
mass of a cell, and then the length $L$ for the cell is chosen
with probability $Z_l^{-1}\exp(E(L)-E(L_T))$. Here the
normalization constant  $Z_l$ is chosen to have the total
probability 1.  In most simulations, we use the following
distribution for the chemical field $c(x)$:
\begin{eqnarray}
\label{cchem1}
     c(x)=\frac{(x-70)^2}{400}.
\end{eqnarray}

\subsection{Monte Carlo simulations versus numerical solutions of the discrete
master equation and the Fokker-Planck equations}

\begin{figure}
~\\
(a)
\\
 \includegraphics[width=0.5\textwidth]{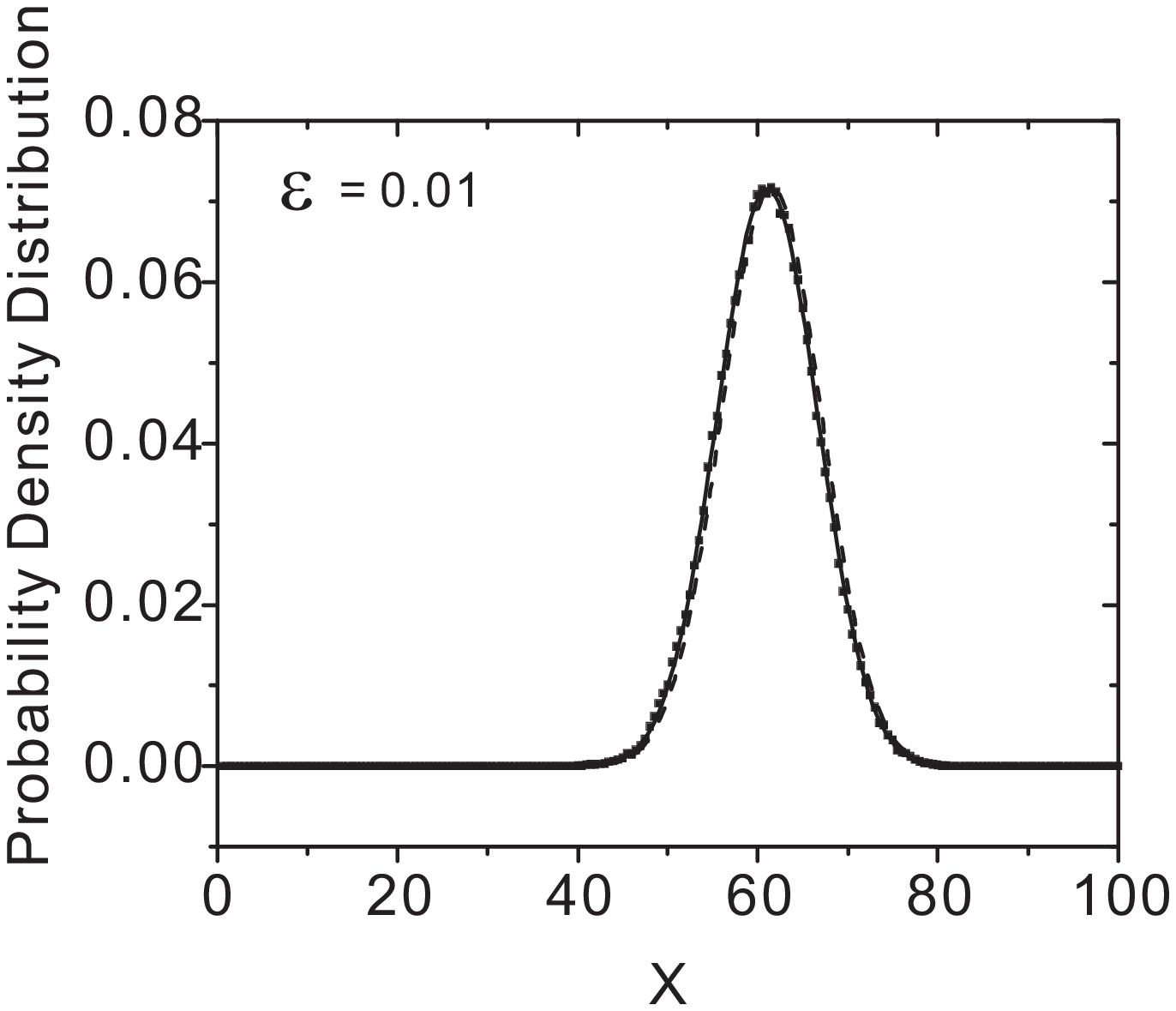}   \vspace{1cm} \\
 (b)
\\
\includegraphics[width=0.5\textwidth]{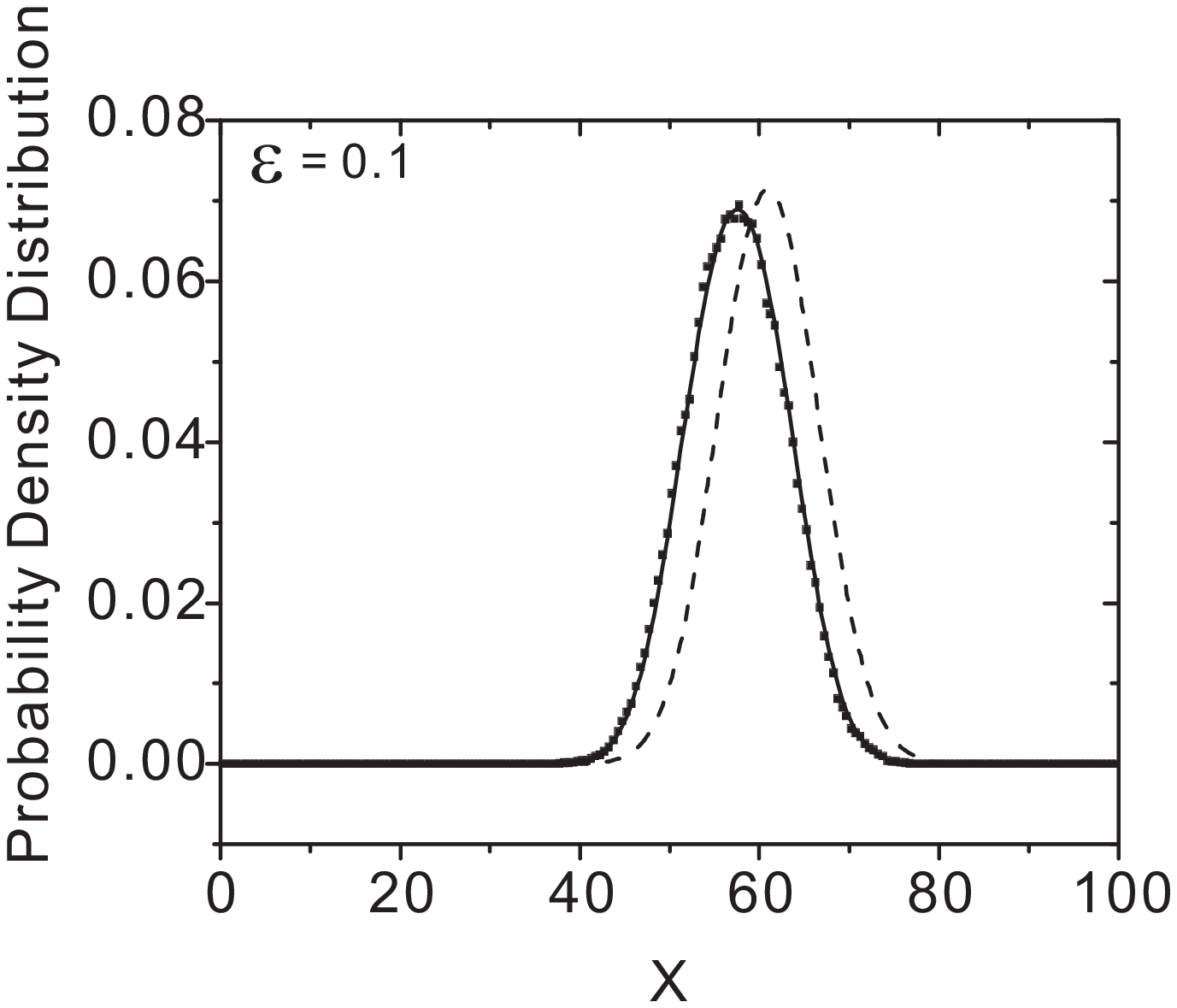}
\caption{Probability densities for Monte Carlo simulations
$p_{\text{cpm}}(x,t)$} (dotted line), $p(x,t)$ for the Master Eq.
$(\ref{pmasterxL1})$ (solid line) and the Fokker-Planck equation
$(\ref{pottscontinuousfull1})$ (dashed line) versus $x$ for
$t=t_{end}.$  (a) $\epsilon=0.01$; (b) $\epsilon=0.1$. The
difference between position of solid curve and a dashed curve is
negligibly small in (a). Number of Monte Carlo simulations is
$N=2\times 10^5$. We used $c(x)$ as given by $(\ref{cchem1}).$
\label{fig:MTvs2D}
\end{figure}

We first compare Monte Carlo simulations with the numerics for the
master equation $(\ref{pmasterxL1})$ and the Fokker-Planck
equation $(\ref{pottscontinuousfull1})$. Simulations of the
Fokker-Planck equation $(\ref{pottscontinuousfull1})$ have been
performed by using a finite-differences scheme. Figure
\ref{fig:MTvs2D} shows the probability density functions for all
three types of simulations.
The difference between the master equation $(\ref{pmasterxL1})$
and the Fokker-Planck equation $(\ref{pottscontinuousfull1})$
simulations is negligibly small  for $\epsilon=0.01$ (Fig.
\ref{fig:MTvs2D}a) but can be clearly seen for $\epsilon=0.1$
(Fig. \ref{fig:MTvs2D}b). We conclude that for $N\to \infty$, the
Monte Carlo simulations converge to the solution of the master
equation $(\ref{pmasterxL1})$ for any $\epsilon$. The rate of
convergence is about $N^{-1/2}$. For small $\epsilon\to 0$, the
solution of the Fokker-Planck equation
$(\ref{pottscontinuousfull1})$ also converges to the solution of
the master equation.

\subsection{Convergence of the probability density function $P(x,L,t)$ to the Boltzmann distribution}
\begin{figure}
\begin{center}
\includegraphics[width = 3.4 in]{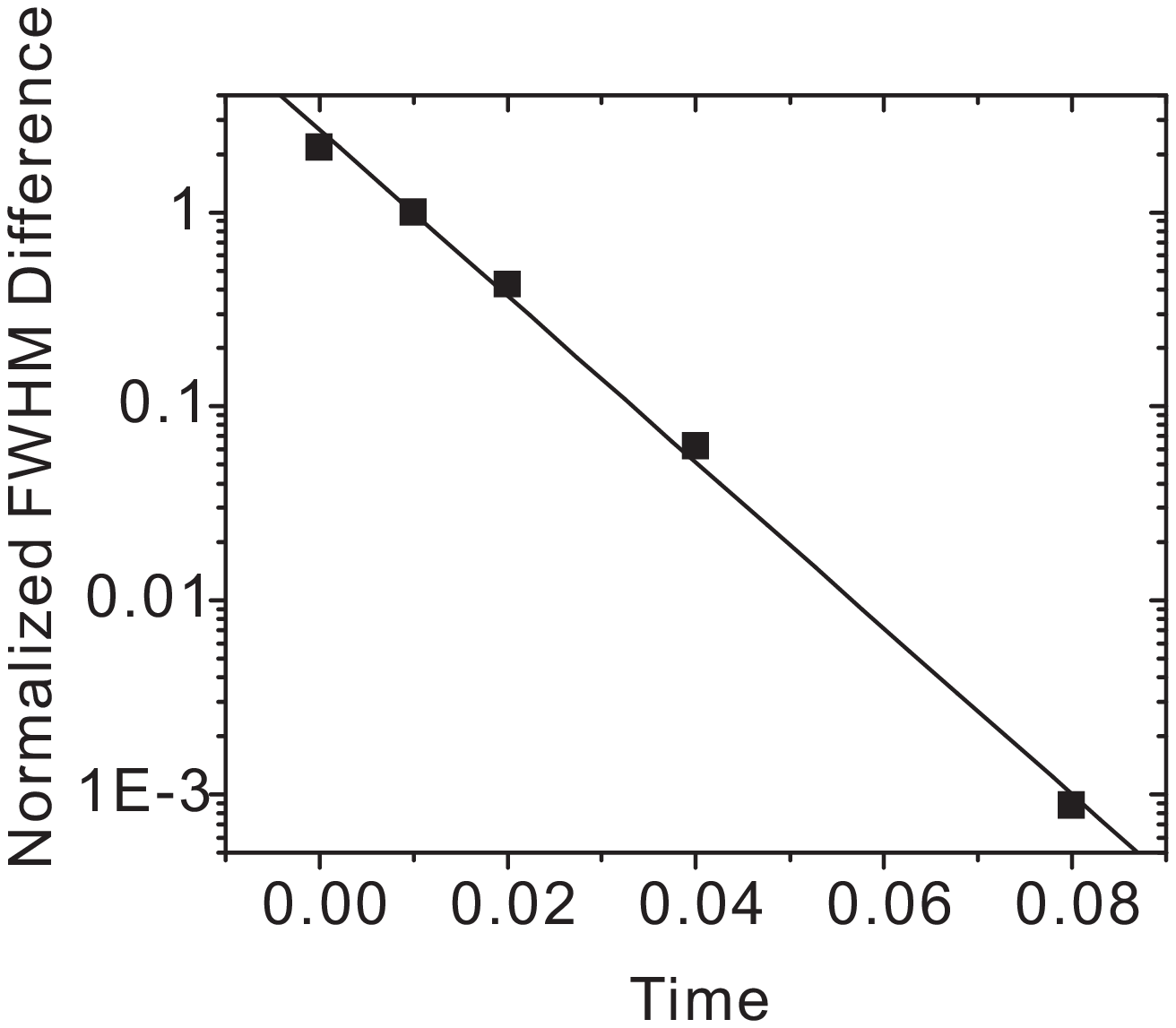}
\caption{Exponential convergence of the Full Width Half Maximum
(FWHM) of $P(x,L,t)$ in $L$ as a function of time for
\textbf{$x=50$}. The vertical axis corresponds to the normalized
difference $\big [ W(t)-W_\beta\big ]/W_\beta$, where $W(t)$ is
the FWHM at time $t$ and $W_\beta$ is the FWHM for the Boltzmann
distribution $(\ref{Pinitial}).$ Solid squares correspond to the
numerical solution of both Eqs. $(\ref{pmasterxL1})$ and
$(\ref{pottscontinuousfull1})$. The solid line is the best linear
fit which gives exponential convergence $e^{-98.55t}$. The same
parameters as in Figure \ref{fig:MTvs2D} are  used here with
$\epsilon=0.01$. } \label{fig:exponentialconvergence}
\end{center}
\end{figure}

To demonstrate    quick convergence of $P(x,L,t)$ to the Boltzmann
distribution $(\ref{Pinitial})$ (as discussed in Section
\ref{continuousCPM}) we solve numerically both the master
equation $(\ref{pmasterxL1})$ and its continuous limit
$(\ref{pottscontinuousfull1})$ with initial conditions $P(x,L,0)$
being different from the Boltzmann distribution
$(\ref{pmasterxL1})$. Namely, we choose initial value $P(x,L,0)$
to be the Bolzmann distribution with different temperature
$\beta_{ini}=1.5$ so that Figure
\ref{fig:exponentialconvergence}  shows convergence of initial
state with temperature $1/\beta_{ini}$ to the quasi-equilibrium
state with temperature $1/\beta=1/15$ used in the Monte Carlo
algorithm. Linear-log  plot in the Figure
\ref{fig:exponentialconvergence} indicates that
convergence is indeed exponential in time with high convergence
rate $\tau_r^{-1}$ ($\tau_r^{-1}=98.55$ for parameters of Fig.
\ref{fig:exponentialconvergence}). By high convergence rate we
mean that the typical convergence  time $\tau_r$ is small compare
with e.g. the diffusion time $x_0^2/D$ in $x$ (see Eq.
$(\ref{pottscontinuousfull1})$). Because of the $x$-dependence of
the chemical field, the convergence rate $\tau_r$ is also
$x$-dependent and a closed analytic expression for it
 is difficult to obtain
 from Eq.
$(\ref{pottscontinuousfull1})$ for general $c(x).$
 However even a simple estimate  $\tau_r^{-1}= 8D\beta\lambda$
of the rate of convergence gives $60$ for parameters of Figure
$\ref{fig:exponentialconvergence}$ which is qualitatively close
to numerical value $98.55$. Here $98.55$ is obtained from the
linear fit presented in Figure $\ref{fig:exponentialconvergence}$.

Also, we observe that if we increase temperature $T$ in Monte
Carlo simulations, so that condition $(\ref{betacond0})$ is not
true any more, then it results in a significant departure from the
Boltzmann distribution $(\ref{Pinitial})$ which confirms the
theoretical results of Section \ref{continuousCPM}.

\subsection{$P(x,L,t)$ vs. $p(x,t)$ simulations \label{Pp}}

The ansatz $(\ref{Pinitial})$ can be used for fast simulations of
solutions of the discrete master equation. Summing up over all
values of $L$ in the master Eq. $(\ref{pmasterxL1})$ and taking
into account $(\ref{Pinitial})$ result in a discrete equation for
the
probability density function $p(x,t)$ 
\begin{eqnarray}\label{pmaster1}
p(x,t+\epsilon^2\triangle t)=\big
[1-T(x+\frac{\epsilon}{2}\triangle x;\,
x,t)-T(x-\frac{\epsilon}{2}\triangle x;\, x,t)\big ]p(x,t)
\nonumber \\
+T(x; \, x-\frac{\epsilon}{2}\triangle
x,t)p(x-\frac{\epsilon}{2}\triangle x,t)+T(x; \,
x+\frac{\epsilon}{2}\triangle x,t)p(x+\frac{\epsilon}{2}\triangle
x,t),
\end{eqnarray}
where $T(x;\, x',t)$ is a transition probability of a change of
position of a center mass from $x'$ to $x$ at time $t$.
Expressions for $T(x;\, x',t)$ are described in the Appendix. They
are calculated only once at the beginning of a simulation which
makes the numerics for discrete Eq. $(\ref{pmaster1})$ very efficient.

We run simulations for the discrete equation $(\ref{pmaster1})$
and the continuous Eq. $(\ref{pottscontinuous2})$ and compared
them with the solutions of the discrete $(\ref{pmasterxL1})$ and
continuous $(\ref{pottscontinuousfull1})$ equations,
respectively. We find, taking into accout Eq. $(\ref{Pinitial})$,
that indeed the differences between these solutions are very
small for the typical values of parameters.

We conclude  that the Monte Carlo simulations of CPM are
equivalent in the limit of large $N$ to the the simulations of the
discrete Eq. $(\ref{pmaster1})$ for any $\epsilon$.

\subsection{Comparison of the continuous model with the CPM}

Below we denote as $p_{cpm}$ both Monte Carlo simulations  and
numerical solutions of Eq. $(\ref{pmaster1})$ and as
$p_{cont}(x,t)$ solutions of $(\ref{pottscontinuous2})$.

\begin{figure}
\includegraphics[width=0.3\textwidth]{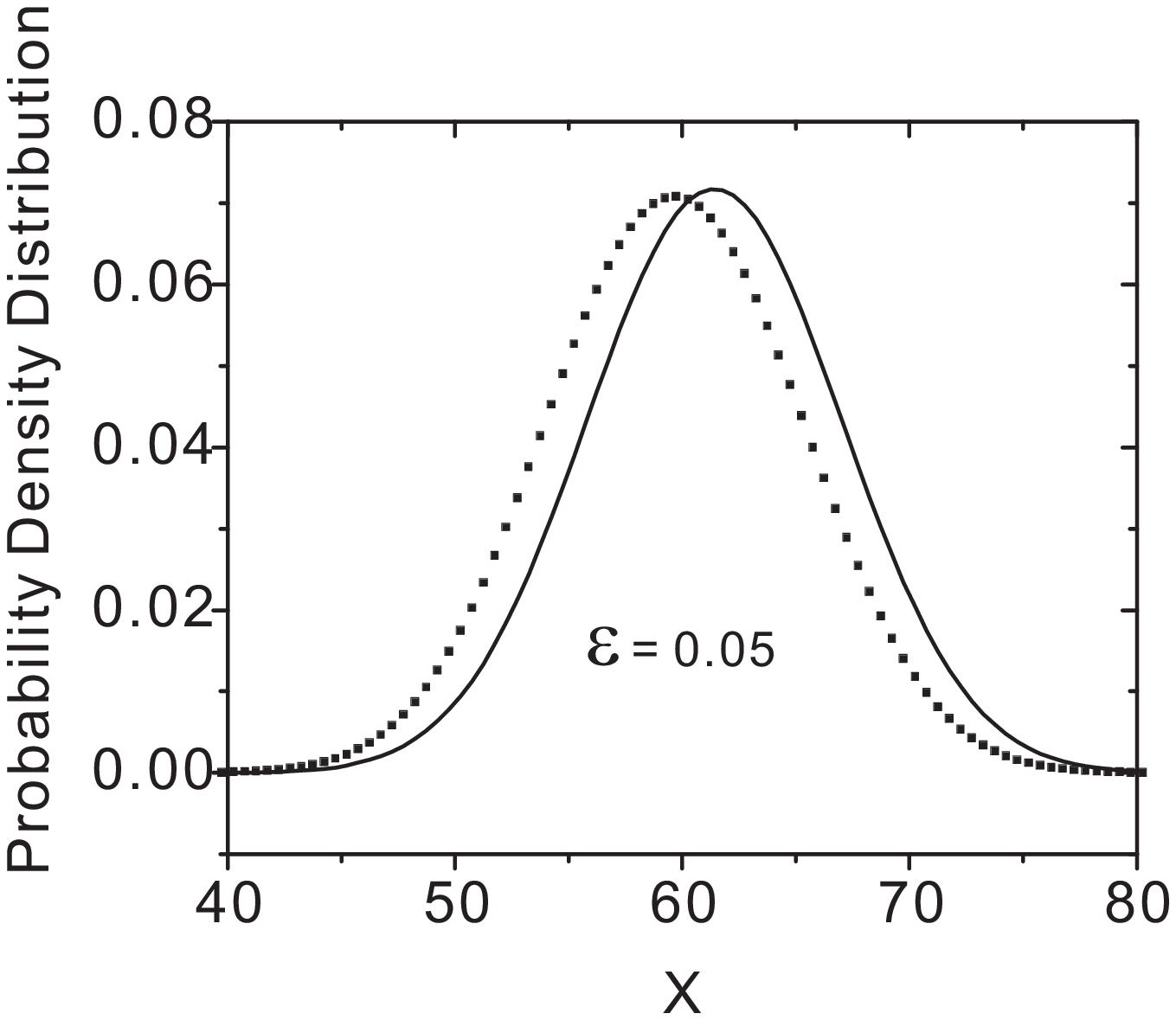}
\includegraphics[width=0.3\textwidth]{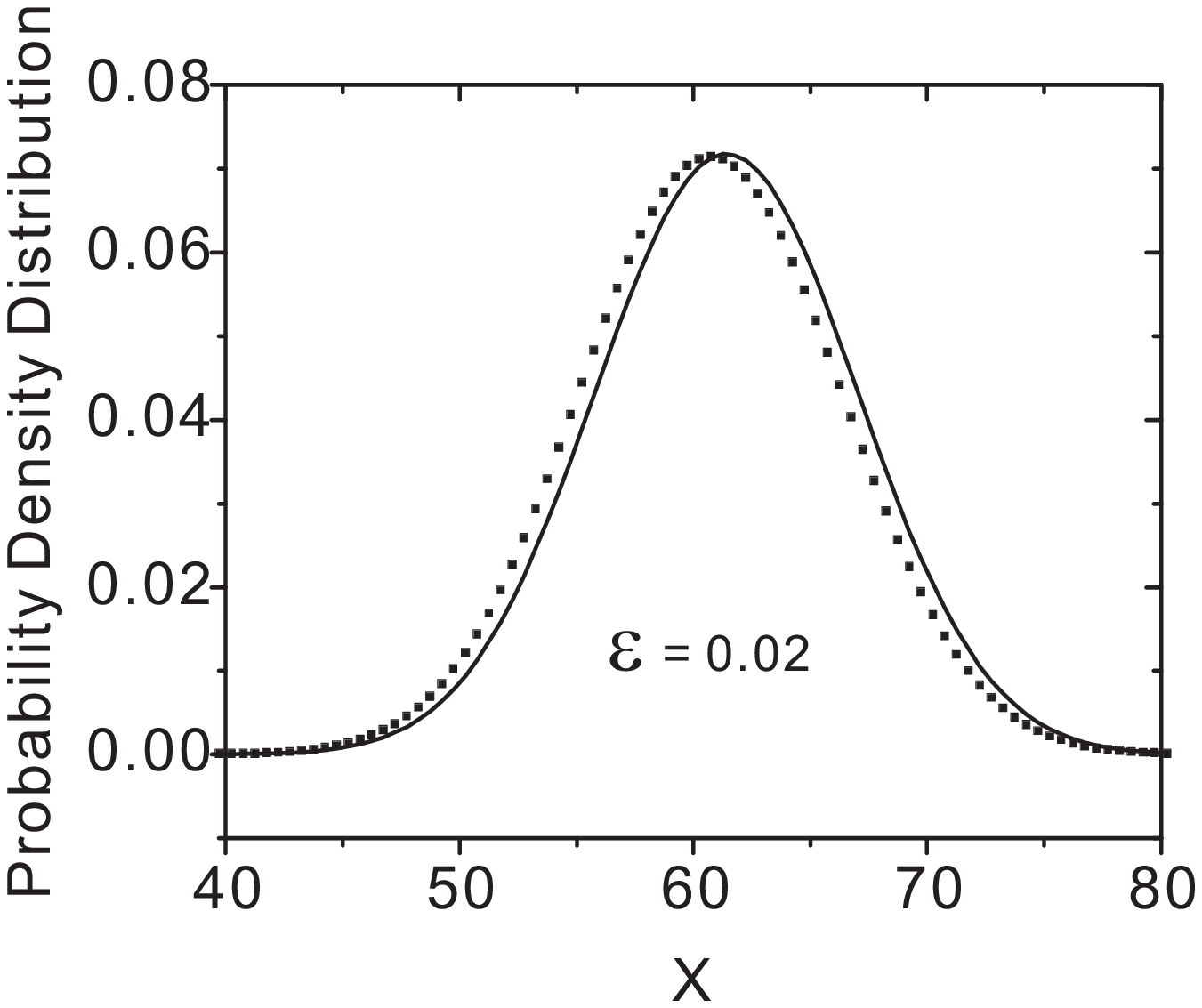}
\includegraphics[width=0.3\textwidth]{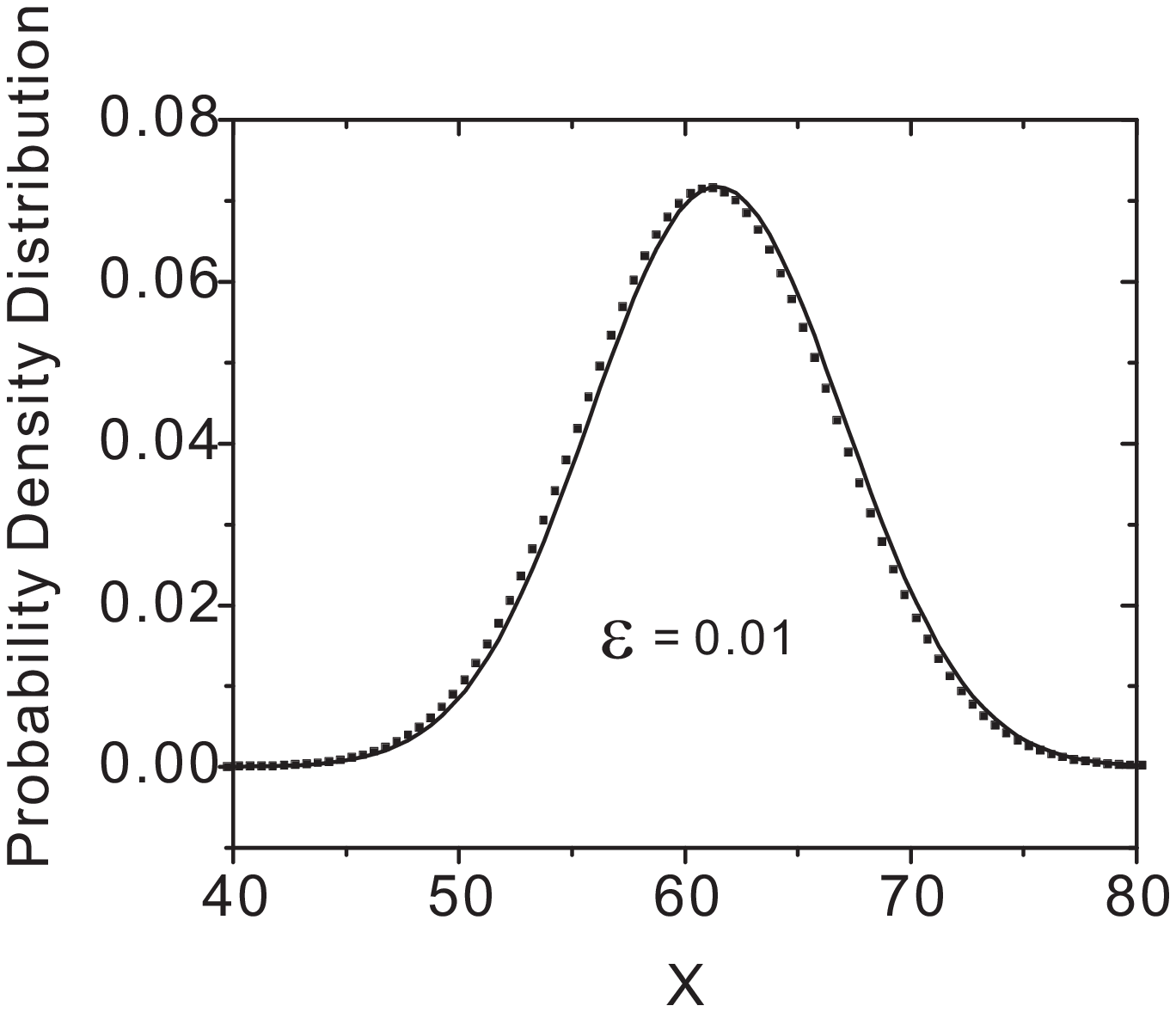}\\
\includegraphics[width=0.3\textwidth]{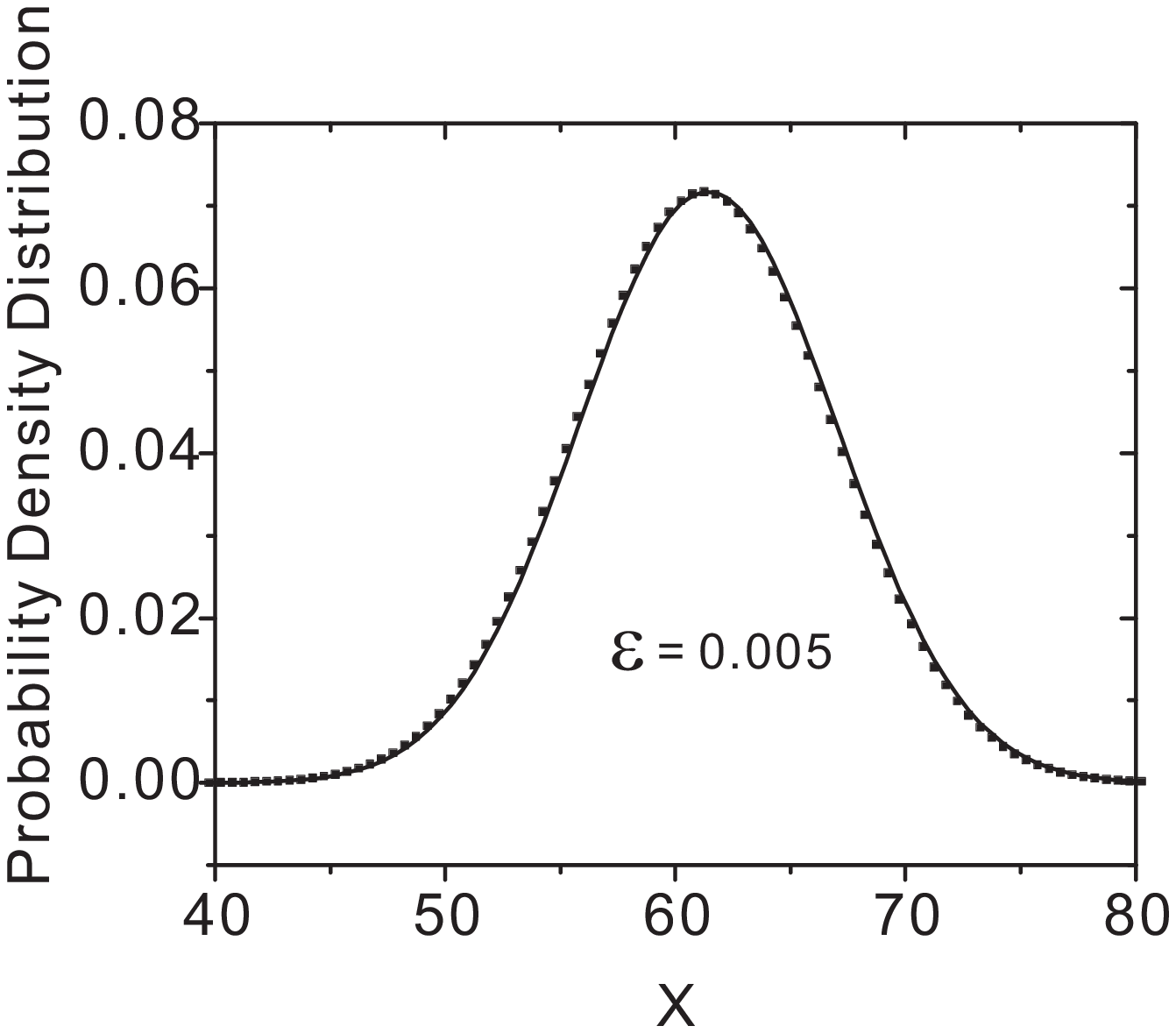}
\includegraphics[width=0.3\textwidth]{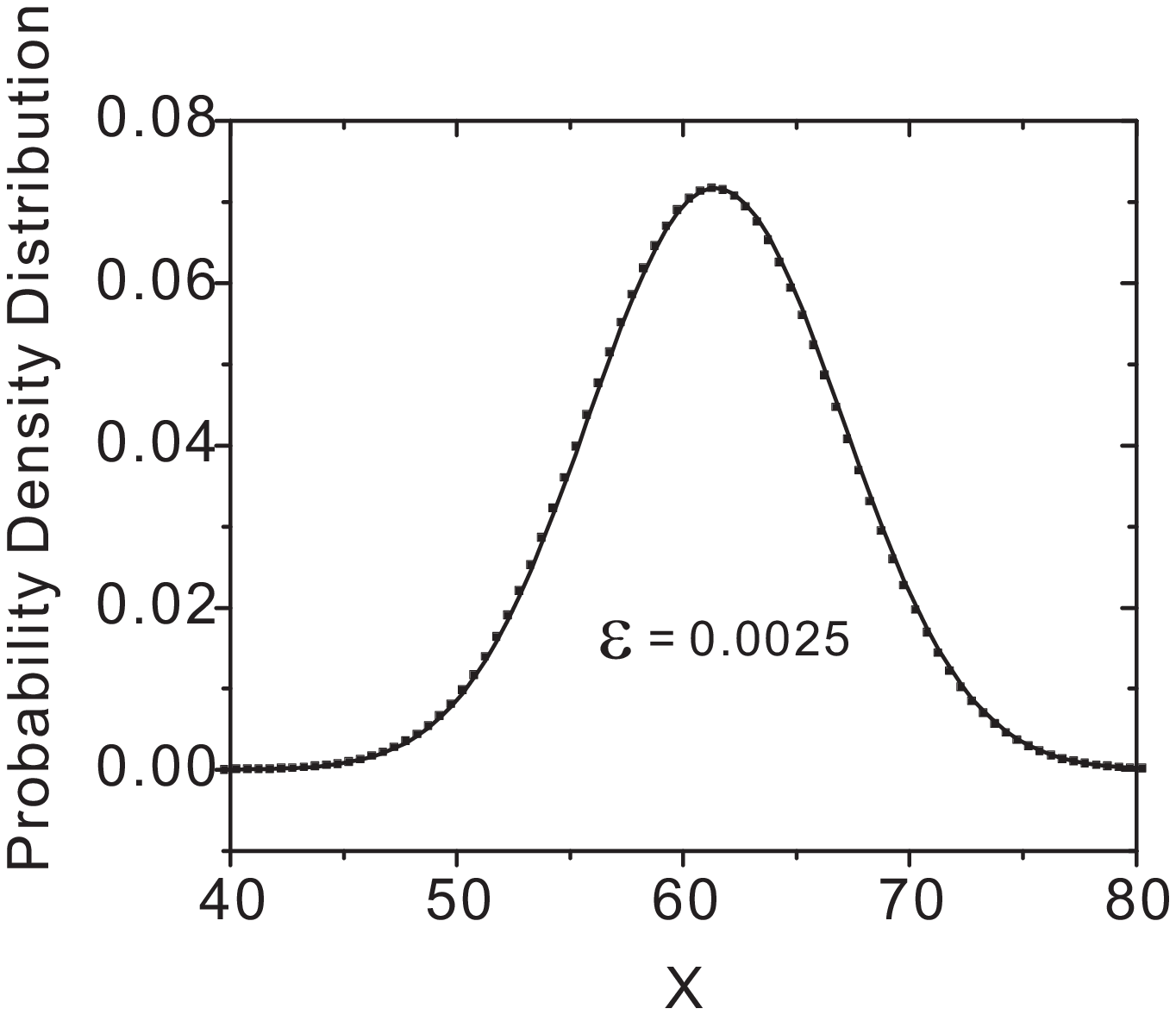}\\
\caption{Plots of $p_{cpm}$ (dotted line) and $p_{cont}(x,t)$
(solid line) as functions of $x$ for a series of decreasing values
of $\epsilon$ at time $t=200$. All other parameters are the same
as in Figure  \ref{fig:MTvs2D}.}\label{eps}
\end{figure}

Figure \ref{eps} shows a series of simulations of the CPM (dotted
line) and numerical solutions of the continuous Eq.
$(\ref{pottscontinuous2})$ (solid line) for different values of
$\epsilon.$
This Figure demonstrates that  in the limit $\epsilon\to 0$, the
solution of the continuous Eq. $(\ref{pottscontinuous2})$ appears
to converge to the cell probability density function of the CPM.

\begin{figure}
\begin{center}
\includegraphics[width = 3.4 in]{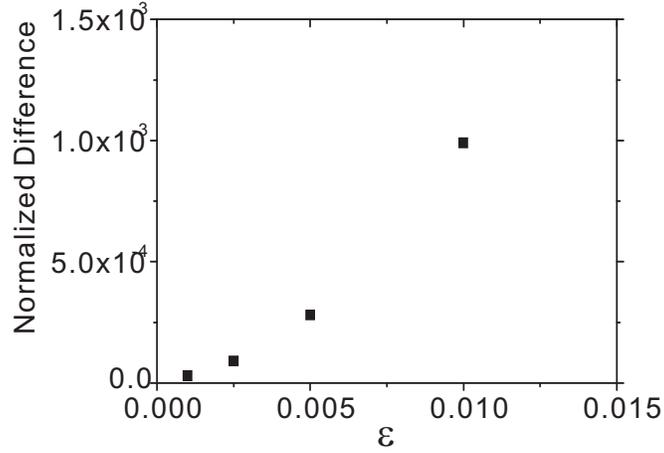}
\caption{Normalized difference between solution of CPM and
continuous Eq. $(\ref{pottscontinuous2})$ for the same parameters
as in Figure \ref{fig:MTvs2D}  as a function of $\epsilon.$
Normalized difference is given by $1-\int
p_{cpm}(x,t)p_{cont}(x,t)dx/\int p_{cont}(x,t)^2dx$ for
$t=t_{end}$.} \label{fig:error}
\end{center}
\end{figure}

\begin{figure}
\includegraphics[width=0.5\textwidth]{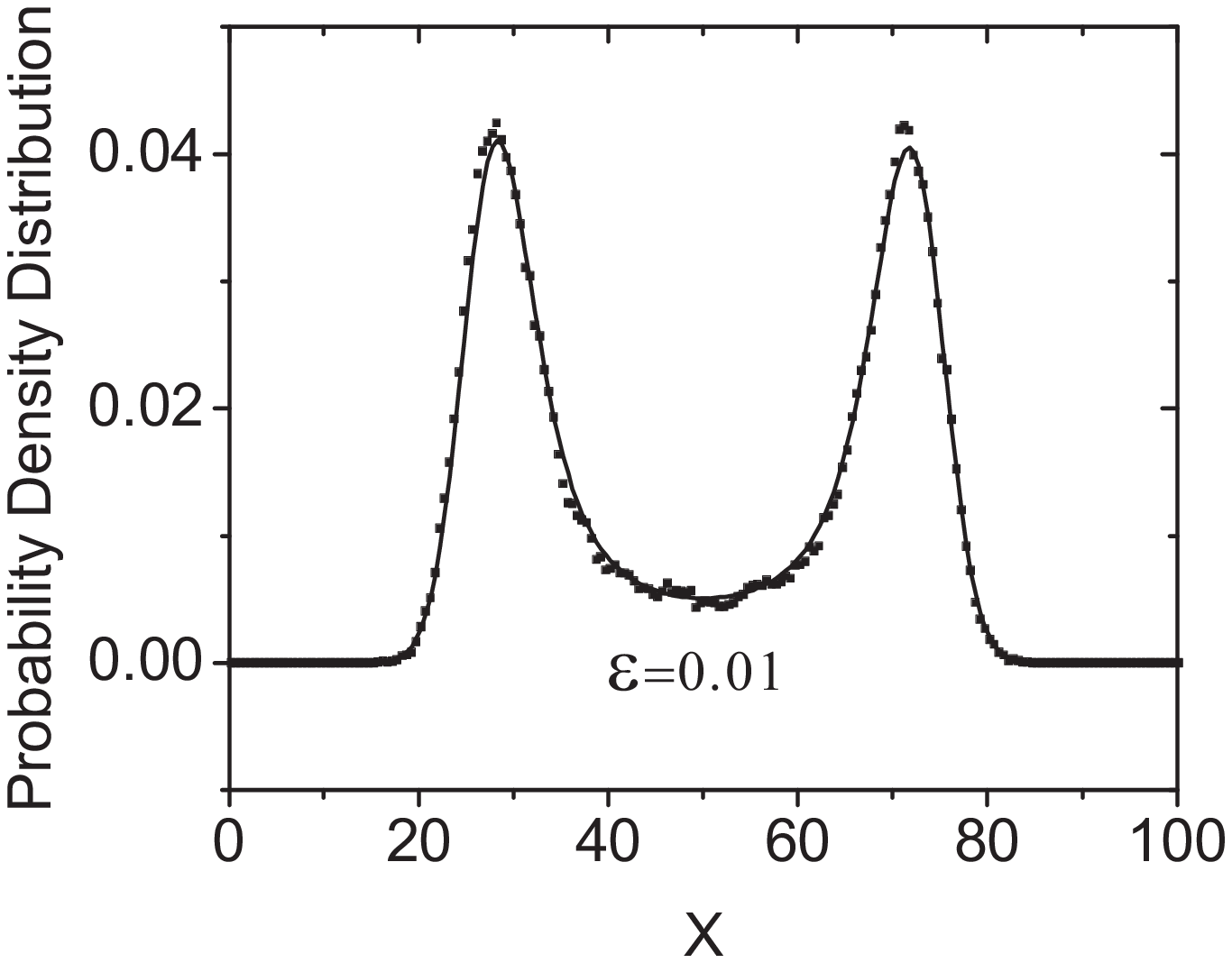}\\
\caption{Typical results of CPM simulations. The same parameters
as in Figure \ref{fig:MTvs2D} are used except that $c(x) =
\cos(4\pi x/100)$, $\epsilon=0.01.$ The same notation for solid,
dashed and dotted curves as in Figure \ref{fig:MTvs2D} is used
here. The difference between position of solid curve and a dashed
curve is again negligibly small.}\label{all}
\end{figure}

Figure \ref{fig:error} shows the normalized difference between
solutions of $(\ref{pottscontinuous2})$ and the CPM. The normalized difference
approaches $0$ as $\epsilon$ decreases.
We also run a series of tests for different forms of the chemical
field $c(x)$ and demonstrate that solutions of the CPM and
continuous Eq. $(\ref{pottscontinuous2})$ are close for small
values of $\e$. Figure~\ref{all} shows a typical result of
numerical simulations for a ``double well'' chemical
concentration $c(x) = \cos(4\pi x/100).$
We conclude that the numerical simulations show excellent agreement
between the CPM and the continuous Eq. $(\ref{pottscontinuous2})$ provided
that the Potts parameters satisfy conditions $(\ref{betacond0})$,
$(\ref{betacond1})$, $(\ref{lambcond1})$, $(\ref{epscond1})$ and
$\epsilon\to 0$, which correspond to the  continuous limit of the CPM.

\section{Conclusions}

In this paper we combine microscopic and macroscopic levels of
description of one dimensional cellular dynamics. The microscopic
level is represented by a one dimensional CPM with chemotaxis and
without cell-cell adhesion term.    We study a continuous
macroscopic limit of our CPM  as the size of Monte-Carlo step is
made small under the assumption that changes in the cell's
position and length  are also small. In this limit, we derive the
Fokker-Planck equation $(\ref{pottscontinuous2})$ for the
probability density function $p(x,t)$ of cells and then further
reduce it to the well-known macroscopic continuous Keller-Segel
model $(\ref{ceq1})$ and $(\ref{pottscontinuousKellerSegel})$ for
the chemotactic aggregation of cells.  All coefficients of the
Keller-Segel model are derived from parameters of the CPM.

We use numerical simulations to test  hierarchy of models and
assumptions which we used to derive continuous equation
$(\ref{pottscontinuous2})$. In particular, we compare  Monte
Carlo simulations with   simulations of both the discrete master
equation $(\ref{pmasterxL1})$ and the Fokker-Planck equation
$(\ref{pottscontinuousfull1})$ for $P(x,L,t)$. We find that, as
expected from our theoretical analysis, all models agree for small
$\epsilon$. Also Monte Carlo simulations agree with   the
solutions of the discrete master equation $(\ref{pmasterxL1})$
for arbitrary $\epsilon$. We verify numerically  that the
probability density function $P(x,L,t)$ quickly converges  to the
Boltzmann distribution $(\ref{Pinitial})$. And finally, we find
that numerical simulations show excellent agreement between Monte
Carlo simulations of CPM and the continuous macroscopic model
$(\ref{pottscontinuous2})$.

We are currently working on extending our results to a 2D case for
modeling chondrogenic patterning in the presence of chemotaxis and
fibronactin production \cite{Kiscowski}.

\section{Acknowledgments}

This work was partially supported by NSF Grant No. IBN-0083653.
Simulations were performed on the Notre Dame Biocomplexity Cluster
supported in part by NSF MRI Grant No. DBI-0420980.

\section{Appendix}

The explicit expressions for the transitional probabilities
$T(x;\, x',t)$ used in Eq. $(\ref{pmaster1})$ can be obtained by
summing over all lengths (or, in other words, over even multiples
of $\epsilon \triangle x$ (if $2x/(\epsilon\triangle x$) is an
even number), and over odd multiples of $\epsilon \triangle x$ (if
$2x/(\epsilon\triangle x$) is an odd number). A change in the
position of the center of mass from $x$ to $x\pm
\frac{\epsilon}{2}\triangle x$ can be made by adding/removing
lattice sites from the left/right end of a cell which results in
\begin{eqnarray}\label{Tcalc1}
T(x; \,x-\frac{\epsilon}{2}\triangle x,t)
=\frac{1}{4Z(x-\frac{\epsilon}{2}\triangle x)}
 \sum\limits_{L=(1+\alpha)\epsilon \triangle x, \,
(3+\alpha)\epsilon \triangle x, \, (5+\alpha)\epsilon \triangle
x,\ldots}
\nonumber
\\
%
\Big \{\exp\big [-\beta \triangle
E_{length}(x-\frac{\epsilon}{2}\triangle x,L-\epsilon\triangle
x)\big ] \Phi\Big (E(x,L)- E(x-\frac{\epsilon}{2}\triangle
x,L-\epsilon\triangle x) \Big )
\nonumber \\
+\exp\big [-\beta \triangle
E_{length}(x-\frac{\epsilon}{2}\triangle x,L+\epsilon\triangle
x)\big ] \Phi\Big (E(x,L)- E(x-\frac{\epsilon}{2}\triangle x,L+\epsilon\triangle x) \Big )\Big \}, \nonumber \\
T(x; \,x+\frac{\epsilon}{2}\triangle x,t)
=\frac{1}{4Z(x+\frac{\epsilon}{2}\triangle x)}
 \sum\limits_{L=(1+\alpha)\epsilon \triangle x, \,
(3+\alpha)\epsilon \triangle x, \, (5+\alpha)\epsilon \triangle
x,\ldots}
\nonumber
\\
%
\Big \{\exp\big [-\beta \triangle
E_{length}(x+\frac{\epsilon}{2}\triangle x,L-\epsilon\triangle
x)\big ] \Phi\Big (E(x,L)- E(x+\frac{\epsilon}{2}\triangle
x,L-\epsilon\triangle x) \Big )
\nonumber \\
+\exp\big [-\beta \triangle
E_{length}(x+\frac{\epsilon}{2}\triangle x,L+\epsilon\triangle
x)\big ] \Phi\Big (E(x,L)- E(x+\frac{\epsilon}{2}\triangle x,L+\epsilon\triangle x) \Big )\Big \}, \nonumber \\
T(x+\frac{\epsilon}{2}\triangle x; \,x,t)=\frac{1}{4Z(x)}
 \sum\limits_{L=(1+\alpha)\epsilon \triangle x, \,
(3+\alpha)\epsilon \triangle x, \, (5+\alpha)\epsilon \triangle
x,\ldots}
\nonumber
\\
%
\Big \{\exp\big [-\beta \triangle E_{length}(x,L)\big ] \Phi\Big
(E(x+\frac{\epsilon}{2}\triangle x,L-\epsilon\triangle x)- E(x,L)
\Big )
\nonumber \\
+\exp\big [-\beta \triangle E_{length}(x,L)\big ] \Phi\Big (E(x+\frac{\epsilon}{2}\triangle x,L+\epsilon\triangle x)- E(x,L) \Big )\Big \}, \nonumber \\
T(x-\frac{\epsilon}{2}\triangle x; \,x,t)=\frac{1}{4Z(x)}
 \sum\limits_{L=(1+\alpha)\epsilon \triangle x, \,
(3+\alpha)\epsilon \triangle x, \, (5+\alpha)\epsilon \triangle
x,\ldots}%
\nonumber
\\
%
\Big \{\exp\big [-\beta \triangle E_{length}(x,L)\big ] \Phi\Big
(E(x-\frac{\epsilon}{2}\triangle x,L-\epsilon\triangle x)- E(x,L)
\Big )
\nonumber \\
+\exp\big [-\beta \triangle E_{length}(x,L)\big ] \Phi\Big (E(x-\frac{\epsilon}{2}\triangle x,L+\epsilon\triangle x)- E(x,L) \Big )\Big \}, \nonumber \\
\nonumber \\
\alpha=1 \ \mbox{for} \ \frac{x}{\epsilon \triangle x}=n, \quad
%
%
\alpha=0 \ \mbox{for} \ \frac{x}{\epsilon \triangle x}=n+1/2,
\quad n\in\mathbb{N}.
\end{eqnarray}
Here the partition function $Z(x)$ is given by $(\ref{Zpart})$.
$Z(x)$ is $x$-dependent in the discrete case considered in this
Appendix. This $x$-dependence is eliminated after going from a
discrete summation in $(\ref{Zpart})$ to an integral (as in Eq.
$(\ref{Zpartcont})$). We evaluate the transitional probabilities
$T(x; \,x\pm\frac{\epsilon}{2}\triangle x,t)$ and
$T(x\pm\frac{\epsilon}{2}\triangle x,t)$ numerically using
$(\ref{Tcalc1})$ for each value of $x$ once at the beginning of
each simulation and then calculate the discrete evolution of Eq.
$(\ref{pmaster1})$.

Notice that in the limit of small $\epsilon \to 0$, the continuous
equation $(\ref{pottscontinuous2})$ can be derived directly from
$(\ref{Zpart}),(\ref{pmaster1})$ and $(\ref{Tcalc1})$. However,
this derivation is more tedious compared with the two-step
derivation in Sections \ref{continuousCPM} and \ref{FokkerPlanck}
where continuous equation $(\ref{pottscontinuousfull1})$ is first
derived and then integrated $(\ref{pottscontinuousfull1})$ over
$L$ which results in Eq. $(\ref{pottscontinuous2})$.

\end{document}